\RequirePackage{fix-cm}
\documentclass[smallextended]{svjour3}       
\smartqed  
\usepackage{graphicx}
\usepackage{amsmath}
\usepackage{amssymb}
\usepackage{pdfpages}
\usepackage[none]{hyphenat}
\begin{document}

\title{Exact solutions of time fractional generalized  Burgers-Fisher Equation using generalized Kudryashov method
}


\author{Ramya Selvaraj$^1$        \and
       V. Swaminathan$^1$ \and A.Durga Devi$^2$ \and K.Krishnakumar$^1$ 
}


\institute{${}^1$ 
             Department of Mathematics,
             Srinivasa Ramanujan Centre, 
              SASTRA Deemed to be University, Kumbakonam 612 001, India.
             \email{rsramyaselvaraj@gmail.com, mvsnew@gmail.com}            \\
       ${}^2$
            Department of Physics,
            Srinivasa Ramanujan Centre, 
             SASTRA Deemed to be University, Kumbakonam 612 001, India.
}
\date{Received: date / Accepted: date}

\maketitle

\begin{abstract}
In this article,  we study the generalised Kudryashov method  for the time fractional generalized Burgers-Fisher equation (GBF). Using traveling wave transformation, the time fractional GBF is transformed to nonlinear ordinary differential equation (ODE). Later, in the nonlinear ODE of time fractional GBF,  the generalized Kudryashov and power series method is applied to get exact solutions. 

\end{abstract}

\section{Introduction}
\hspace{\parindent} The fractional differential equation (FDE) plays a vital role in many branches of science and engineering \cite{kilbas,juma,eslami,podlubny,huan}. FDE has many applications in the field of magnetism, sound waves propagation in rigid porous materials, cardiac tissue electrode interface, theory of viscoelasticity,  fluid mechanics,  lateral and longitudinal control of autonomous vehicles, ultrasonic wave propagation in human cancellous bone,  wave propagation in viscoelastic horn, heat transfer, RLC electric circuit  and so on.
    
    In recent years, for solving time fractional differential equations,
    many researchers have proposed powerful techniques to get an exact solution such as, the sine-cosine method \cite{Bekir,Mirzazadeh}, $G'/G$ expansion method \cite{Ilhan}, the Exp-function method \cite{He}, the tanh method \cite{Wazwaz}, the sub equation method \cite{Alzaidy,Guo}, the improved $G'/G$ expansion method \cite{AkbarNaher}, the invariant subspace method, the generalized Riccati equation method \cite{Naher}, the modified Kudryashov method \cite{bibi,Demiray,Gepreel,bejarbanehhosseini,mayelihosseiniansari,mayelihosseini,kaplan,koparankaplan,korkmazhosseini,Mahmud}  and so on.
   
   Kudryashov method was introduced by Kudryashov \cite{kudrya} for reliable treatment of nonlinear wave equations. For both integer and fractional order, this method is widely used by many researchers such as \cite{bulutbaskonus,ryabov,egemisirli1,egemisirli2,bulutpandir,zayed}. In this work, we apply Kudryashov method to the time fractional generalized Burgers-Fisher equation.
   
   A nonlinear equation which is the combination of reaction, convection and diffusion mechanism is called Burgers-Fisher equation. In the nonlinear equation, the properties of convective phenomenon from Burgers and  diffusion transport as well as reaction kind of characteristics from Fisher are used.  The generalized Burgers-Fisher (GBF) equation  is used in the field of fluid dynamics. It has also been found in some applications such as gas dynamics, heat conduction, elasticity and so on.
   
The aim of this work is organized as follows: In section 2, the algorithmic procedure of generalized Kudryashov method (GKM) is proposed. In section 3, we apply the GKM to find the exact solutions of time fractional generalized Burgers-Fisher equation. Power series method is applied to find the explicit solution of the time fractional GBF equation in section 4 and  section 5 ends with conclusion.

\section{Algorithm of the generalized Kudryashov Method}
 \hspace{\parindent} Let us consider a function $u$ of two real variables, space $x$ and time $t$, then the NLPDE  with fractional order can be written in the form of
\begin{equation} \label{FPDE}
  P(u,D^\alpha_tu,u_x,u_{xx},\ldots)=0
\end{equation}
We illustrate the main steps of generalized Kudryashov method as follows:\\
    In the first step, we obtain the traveling wave solution of Eq.(\ref{FPDE}) of the form:
    \begin{equation}\label{travelingwave}
    u(x,t)=u(\xi),    \;   \xi=kx-\frac{\lambda t^\alpha}{\Gamma[1+\alpha]}, 
\end{equation}
where $k$ and $\lambda$ are arbitrary constants. As a result of this, we obtain a nonlinear ODE in the following form:
\begin{equation} \label{conversionODE}
    N(u,u^\prime,u^{\prime \prime},u^{\prime \prime \prime}, \ldots)=0,
\end{equation}
where the prime indicates differentiation with respect to $\xi$.\\
In the second step, the exact solutions of the nonlinear ODE can be written in the following form:
\begin{equation} \label{divisionodpandq}
u(\xi)=\frac{\sum^K_{i=0}p_iR^i(\xi)}{\sum^N_{j=0}q_jR^j(\xi)}=\frac{P[R(\xi)]}{Q[R(\xi)]},
\end{equation}
where $R$ is $\frac{1}{(1\pm e^\xi)}$. Then 
\begin{equation}
    R_{\xi}=R^2-R.
\end{equation}
Then we obtain
\begin{eqnarray}
u^\prime(\xi)&=&\frac{P^\prime R^\prime Q-PQ^\prime R^\prime}{Q^2}\nonumber\\ 
&=&R^\prime \left[\frac{P^\prime Q-PQ^\prime}{Q^2}\right] \nonumber\\
   &=&(R^2-R)\left[\frac{P^\prime Q-PQ^\prime}{Q^2}\right],\\
 u^{\prime\prime}(\xi)&=&\displaystyle \left[\frac{(R^2-R)}{Q^2}\right]\Bigg[(2R-1)(P^\prime Q-PQ^\prime)\nonumber \\ && +\left(\frac{R^2-R}{Q}\right) [Q(P^{\prime\prime}Q-PQ^{\prime \prime})-2Q^\prime P^\prime P+2P(Q^\prime)^2]\Bigg],\\
 u^{\prime\prime\prime}(\xi)&=&(R^2-R)^3 \Bigg[((P^{\prime\prime \prime}Q-PQ^{\prime\prime \prime}-3P^{\prime\prime}Q^\prime -3Q^{\prime\prime}P^\prime)Q \nonumber \\
 && +6Q(PQ^{\prime\prime}+Q^\prime P^\prime))(Q^3)^{-1}-\frac{6P(Q^\prime)^3}{Q^4}\Bigg] \nonumber \\
 &&+3(R^2-R)^2 (2R-1) \Bigg[\frac{Q(P^{\prime\prime}Q-PQ^{\prime \prime})-2Q^\prime P^\prime P+2P(Q^\prime)^2}{Q^3}\Bigg]
 \nonumber \\
 && +(R^2-R)(6R^2-6R+1)\Bigg[\frac{P^\prime Q-PQ^\prime}{Q^2}\Bigg]
\end{eqnarray}
and so on.\\
  In the third step, we can explain the nonlinear ODE in the form:
   \begin{equation} \label{kudryexactsolution}
       u(\xi)=\frac{p_0+p_1 R+p_2 R^2+\ldots+p_K R^K+\ldots}{q_0+q_1 R+q_2 R^2+\ldots+q_N R^N+\ldots}
   \end{equation}
   By balancing the highest order nonlinear terms  and the highest order derivatives of $u(\xi)$ in Eq.(\ref{conversionODE}) we get some values of $N$ and $K$.\\
    In the fourth step, to calculate the constant $p_i$ and $q_j$, substituting Eq.(\ref{divisionodpandq}) into Eq.(\ref{conversionODE}), it provides a polynomial $R(\xi)$. Establishing the coefficients to zero provides a system of algebraic equations. Solving the system of equations, we can find out the constant. By this way, the exact solution of Eq.(\ref{FPDE}) can be found.

   \section{Applications}
   In this section, we apply the generalized Kudryashov method to the time fractional generalized Burgers-Fisher equation.
    \subsection{Time fractional generalized Burgers-Fisher Equation}
    Let us consider the time fractional generalized Burgers-Fisher equation 
\begin{equation} \label{GBF}
u^\alpha _t+\beta u^\delta u_x-u_{xx}=\gamma u(1-u^\delta) 
\end{equation}
where $0<\alpha \leq 1$, $\alpha$ is the order of the fractional time derivative and $\beta,\gamma,\delta$ are arbitrary constants.
    Now substituting Eq.(\ref{travelingwave}) in Eq.(\ref{GBF}), we obtain
    \begin{equation} \label{tvwreductioneqn}
        k^2 u^{\prime\prime}+(\lambda-k\beta u^{\delta})u^{\prime}+\gamma u(1-u^{\delta})=0
    \end{equation}
Applying the folding transformation
    \begin{equation}
        u(\xi)=v^{\frac{1}{\delta}}(\xi),
    \end{equation}
    we obtain the following equation which is similar to  the  most general form of second order nonlinear oscillator equation \cite{pradeep,mohanasuba} with many arbitrary parameters. With some restrictions on the parameters, they found new integrable equations and further it was discussed by \cite{kkr}. 
\begin{equation}\label{nonlinearode}
   k^2\delta vv^{\prime \prime}+ k^2(1-\delta){v^\prime}^2+( \lambda -k \beta v)\delta v {v^\prime} +\gamma \delta^2 (1-v)v^2=0.
\end{equation}
Balancing the highest power of nonlinear terms of $v^{\prime \prime}$ and $v^3$, then
\begin{equation}
    2K-2N+2=3K-3M \Rightarrow K=N+2.
\end{equation}
Let us choose $N=1,K=3$, then
\begin{equation} \label{riccatifraction}
    v(\xi)=\frac{p_0+p_1R+p_2 R^2+p_3 R^3}{q_0+q_1R},
\end{equation}
\begin{eqnarray}
v^\prime(\xi)&=&(R^2-R) \nonumber \\
&& \times \Bigg[\frac{(p_1+2p_2R+3p_3R^2)(q_0+q_1R)-q_1(p_0+p_1R+p_2 R^2+p_3 R^3)}{(q_0+q_1R)^2}\Bigg], \nonumber\\
v^{\prime \prime}(\xi)&=&\left(\frac{R^2-R}{q_0+q_1 R^2}\right)\Bigg[(2R-1)((p_1+2p_2R+3p_3R^2)(q_0+q_1R^2)\nonumber \\
&&-q_1(p_0+p_1R+p_2 R^2+p_3 R^3))\nonumber \\
&&+\left(\frac{R^2-R}{q_0+q_1R}\right) \Big[(q_0+q_1R)(2p_2+6p_3R)(q_0+q_1R)\nonumber\\
&&-2q_1(p_1+2p_2R+3p_3R^2)(q_0+q_1R) \nonumber \\
&& +2(p_0+p_1R+p_2 R^2+p_3 R^3)q^2_1\Big]\Bigg].
\end{eqnarray}

 The exact solutions of Eq.(\ref{tvwreductioneqn}) are obtained as follows and there are two cases to be considered.
\case 

   For the choices of 

    \[p_2=p_3=0, \;q_0=p_0,\; q_1=0,\; 
    p_1=\frac{k(1+\delta)p_0}{\beta \delta},\] 
    \[\lambda={k^2+k\beta -\gamma \delta}, \;k=\frac{-\beta \delta}{(1+\delta)}\]
and substitutes in Eq.(\ref{riccatifraction}), we get the equation
\begin{eqnarray}
    &&-\beta^2 \delta vv^{\prime \prime}+\beta^2(\delta-1){v^\prime}^2+(\beta^2+\gamma (1+\delta)^2) vv^{\prime}-\gamma(1+\delta)^2 v^3 \nonumber\\
   && -(1+\delta)(\gamma+\gamma \delta+\beta^2 v^\prime)v^2=0.
\end{eqnarray}
The solutions are,
\begin{equation}
 v_1(x,t)=1-\Bigg[\frac{1}{1+ e^{\Big( \frac{\beta^2 \delta t^\alpha + \gamma \delta(1+\delta)^2 t^\delta - \beta \delta (1+\delta)x \Gamma(1+\alpha)}{(1+\delta)^2 \Gamma(1+\alpha)}\Big)}}\Bigg]
\end{equation}
and
\begin{equation}
 v_2(x,t)=1+\Bigg[\frac{1}{1- e^{\Big(\frac{\beta^2 \delta t^\alpha + \gamma \delta(1+\delta)^2 t^\delta - \beta \delta (1+\delta)x \Gamma(1+\alpha)}{(1+\delta)^2 \Gamma(1+\alpha)}\Big)}}\Bigg].
\end{equation}
  Then the exact solutions are,
  \begin{equation} \label{case1exact}
   u_1(x,t)=\Bigg[1-\frac{sec h \Big(\frac{\beta^2 \delta t^\alpha + \gamma \delta(1+\delta)^2 t^\delta - \beta \delta (1+\delta) x \Gamma(1+\alpha)}{2(1+\delta)^2 \Gamma(1+\alpha)} \Big)}{2 e^{\Big(\frac{\beta^2 \delta t^\alpha + \gamma \delta(1+\delta)^2 t^\delta - \beta \delta (1+\delta) x \Gamma(1+\alpha)}{2(1+\delta)^2 \Gamma(1+\alpha)}\Big)}}\Bigg]^\frac{1}{\delta} 
   \end{equation}
   and
   \begin{equation}\label{case1exact1}
   u_2(x,t)=\Bigg[1+\frac{cosec h \Big(\frac{\beta^2 \delta t^\alpha + \gamma \delta(1+\delta)^2 t^\delta - \beta \delta (1+\delta) x \Gamma(1+\alpha)}{2(1+\delta)^2 \Gamma(1+\alpha)} \Big)}{2 e^{\Big(\frac{\beta^2 \delta t^\alpha + \gamma \delta(1+\delta)^2 t^\delta - \beta \delta (1+\delta) x \Gamma(1+\alpha)}{2(1+\delta)^2 \Gamma(1+\alpha)}\Big)}}\Bigg]^\frac{1}{\delta}.
  \end{equation} 
  
 \case 
Considering

    \[p_2=p_3=0,\; q_0=p_0,\;q_1=0,\;
    p_1=\frac{k(1+\delta)p_0}{\beta \delta},\]\\
     \[\lambda={k^2+k\beta -\gamma \delta},\; \gamma=\frac{-k^2(\delta-1)}{\delta^2}\]
and substitutes in Eq.(\ref{riccatifraction}), we get the equation
\begin{equation}
    k vv^{\prime\prime}-k \beta v^\prime v^2+\Big(k+{\beta}\Big)vv^\prime =0.
\end{equation}
The solutions are
\begin{equation}
 v_1(x,t)=1+\frac{2k}{\beta \Big(1+e^{\frac{k x \Gamma(1+\alpha) - (k^2 + k \beta)t^\alpha}{\Gamma(1+\alpha)}}\Big)}
 \end{equation}
 and
 \begin{equation}
 v_2(x,t)=1+\frac{2k}{\beta \Big(1-e^{\frac{k x \Gamma(1+\alpha) - (k^2 + k \beta)t^\alpha}{\Gamma(1+\alpha)}}\Big)}.
\end{equation}
Then the exact solutions are,
\begin{equation} \label{case2exact}
   u_1(x,t)=\displaystyle \Bigg[1+\frac{k sec h \Big(\frac{k x \Gamma(1+\alpha) - (k^2 + k \beta)t^\alpha}{2\Gamma(1+\alpha)}\Big)  }{\beta e^{\Big(\frac{k x \Gamma(1+\alpha) - (k^2 + k \beta)t^\alpha}{2\Gamma(1+\alpha)}\Big) }}\Bigg]^ \frac{1}{\delta} 
   \end{equation}
   and
   \begin{equation} \label{case2exact2}
   u_2(x,t)=\displaystyle\Bigg[1-\frac{k cosec h \Big(\frac{k x \Gamma(1+\alpha) - (k^2 + k \beta)t^\alpha}{2\Gamma(1+\alpha)}\Big)  }{\beta e^{\Big(\frac{k x \Gamma(1+\alpha) - (k^2 + k \beta)t^\alpha}{2\Gamma(1+\alpha)}\Big)} }\Bigg]^ \frac{1}{\delta},
\end{equation}
where $\delta=1$.

\section{Exact Power Series Solutions}
Based on the power series method \cite{Galaktionov,beck,asmar} and symbolic computations \cite{Qiao,Tang,Tian2,Tian1,Tian3}, we  construct the exact power series solutions of Eq.(\ref{GBF}) which is differentiable. First we make use of a transformation
\begin{equation}\label{travwave}
    u(x,t)=u(\eta),       \eta=k x-\frac{\lambda t^\alpha}{\Gamma[1+\alpha]}, 
\end{equation}
where $k$ and $\lambda$ are arbitrary constants with k, $\lambda \neq 0$. Substituting Eq.(\ref{travwave}) into Eq.(\ref{GBF}), then we can get the nonlinear ODE 
\begin{equation}\label{nonlinearode1}
    k^2\delta vv^{\prime \prime}+k^2(1-\delta){v^\prime}^2+( \lambda -k \beta v)\delta v {v^\prime} +\gamma \delta^2 (1-v)v^2=0.
\end{equation}
We suppose that Eq.(\ref{nonlinearode}) has the following solution:
\begin{equation}\label{newpowerseries}
    v(\xi)=\sum^\infty_{n=0} q_n \xi^n,
\end{equation}
where $q_n (n=0,1,2,...)$ are constants. Then, we have 
\begin{eqnarray}\label{newpowerseries1}
v^{\prime} (\xi)=\sum^\infty_{n=0}(n+1) q_{n+1} \xi^n,
\end{eqnarray}
\begin{eqnarray}\label{newpowerseries2}
v^{\prime \prime} (\xi)=\sum^\infty_{n=0}(n+1)(n+2) q_{n+2} \xi^n.
\end{eqnarray}
Substituting Eq.(\ref{newpowerseries}),Eq.(\ref{newpowerseries1}) and Eq.(\ref{newpowerseries2}) into Eq.(\ref{nonlinearode1}) and solving, we get 
\begin{equation}\label{qtwo}
    q_2=\frac{1}{2}\Big[\Big(\frac{\delta-1}{\delta}\Big)\frac{q_1^2}{q_0}-\frac{\lambda q_1}{k^2}+\frac{\beta q_0 q_1}{k}-\frac{\gamma \delta q_0}{k^2}+\frac{\gamma \delta q_0^2}{k^2}\Big], 
\end{equation}
when $n=0$.
When $n\geq1$, we obtain
\begin{eqnarray}\label{qn}
q_{n+2}&=&\frac{1}{(n+2)(n+1)}\Big[\Big(\frac{\delta-1}{\delta}\Big)\frac{(n+1){q^2_{n+1}}}{q_n}-\frac{\lambda (n+1)q_{n+1}}{k^2}\nonumber\\
&&+\frac{\beta (n+1)q_n q_{n+1}}{k}-\frac{\gamma \delta q_n}{k^2}+\frac{\gamma \delta q_n^2}{k^2}\Big].
\end{eqnarray}

It is easy to prove the convergence of the power series
Eq.(\ref{newpowerseries}) with the coefficients given in Eq.(\ref{qtwo}) and Eq.(\ref{qn}). Therefore this power series solution of Eq.(\ref{newpowerseries}) is an exact analytic solution.\\
 Hence, the power series solution of Eq.(\ref{newpowerseries}) can be written as follows:

\begin{eqnarray}
v(\xi)&=&q_0+q_1(\xi)+q_2(\xi^2)+\sum^\infty_{n=1}q_{n+2}\xi^{n+2}
\nonumber \\
&=&q_0+q_1(\xi)+\frac{1}{2}\Big[\Big(\frac{\delta-1}{\delta}\Big)\frac{q_1^2}{q_0}-\frac{\lambda q_1}{k^2}+\frac{\beta q_0 q_1}{k}-\frac{\gamma \delta q_0}{k^2}+\frac{\gamma \delta q_0^2}{k^2}\Big]\xi^2   \nonumber\\
&&+\sum^\infty_{n=1}\frac{1}{(n+2)(n+1)}\Big[\Big(\frac{\delta-1}{\delta}\Big)\frac{(n+1){q^2_{n+1}}}{q_n}-\frac{\lambda (n+1)q_{n+1}}{k^2}\nonumber\\&&+\frac{\beta (n+1)q_n q_{n+1}}{k}-\frac{\gamma \delta q_n}{k^2}+\frac{\gamma \delta q_n^2}{k^2}\Big]\xi^{n+2}.
\end{eqnarray}

\section{Conclusion}
In this article,  the time fractional generalized Burgers-Fisher equation have been transformed into a nonlinear ordinary differential equation by using folding transformation. The resultant equation  is similar to the most familiar general form of second order nonlinear oscillator equation  with some restrictions on the parameters. Then we applied generalized Kudryashov method and also power series method to the resultant equation for finding  exact solutions. Plots are given for the exact solutions which show the dynamics of solutions with suitable parametric choices. 

\begin{acknowledgements}
The authors thank the Department of Science and Technology-Fund Improvement of S\&T Infrastructure in Universities and Higher Educational Institutions Government of India (SR/FST/MSI-107/2015) for carrying out this research work.
\end{acknowledgements}




\end{document}